\documentclass[11pt,a4paper,english,nofootinbib]{revtex4}
\usepackage{lmodern}

\usepackage[T1]{fontenc}
\usepackage[latin9]{inputenc}
\usepackage{color}
\usepackage{babel}
\usepackage{amsmath}
\usepackage{amssymb}
\usepackage{esint}
\usepackage[unicode=true,pdfusetitle,
 bookmarks=true,bookmarksnumbered=false,bookmarksopen=false,
 breaklinks=false,pdfborder={0 0 1},backref=false,colorlinks=false]
 {hyperref}

\makeatletter

\pdfpageheight\paperheight
\pdfpagewidth\paperwidth

\@ifundefined{textcolor}{}
{%
 \definecolor{BLACK}{gray}{0}
 \definecolor{WHITE}{gray}{1}
 \definecolor{RED}{rgb}{1,0,0}
 \definecolor{GREEN}{rgb}{0,1,0}
 \definecolor{BLUE}{rgb}{0,0,1}
 \definecolor{CYAN}{cmyk}{1,0,0,0}
 \definecolor{MAGENTA}{cmyk}{0,1,0,0}
 \definecolor{YELLOW}{cmyk}{0,0,1,0}
}

\usepackage{latexsym}\usepackage{bm}

\makeatother

\definecolor{green}{RGB}{0, 180, 0}
\definecolor{cyan}{RGB}{0, 180, 180}
\definecolor{yellow}{RGB}{211,211,0}

\makeatother

\begin{document}

\title{Near Horizon Symmetry and Entropy Formula for Kerr-Newman (A)dS Black Holes }

\author{Mohammad Reza Setare}
\email{rezakord@ipm.ir}

\affiliation{{Department of Science, Campus of Bijar, University of Kurdistan, Bijar, Iran }}

\author{Hamed Adami}
\email{hamed.adami@yahoo.com}

\affiliation{{Research Institute for Astronomy and Astrophysics of Maragha (RIAAM),
P.O. Box 55134-441, Maragha, Iran }}

\begin{abstract}
In this paper we provide the first non-trivial evidence for universality of the entropy formula $4\pi J_{0}^{+}J_{0}^{-}$ beyond pure Einstein gravity in 4-dimensions. We consider the Einstein-Maxwell theory in the presence of cosmological constant, then write near horizon metric of the Kerr-Newman (A)dS black hole in the Gaussian null coordinate system. We consider near horizon fall-off conditions for metric and $U(1)$ gauge field. We find asymptotic combined symmetry generator, consists of diffeomorphism and $U(1)$ gauge transformation, so that it preserves fall-off conditions. Consequently, we find supertranslation, supperrotation and multiple-charge modes and then we show that the entropy formula is held for the Kerr-Newman (A)dS black hole. Supperrotation modes suffer from a problem. By introducing new combined symmetry generator, we cure that problem.
\end{abstract}

\maketitle

\section{Introduction}\label{S.I}
It seems that any valid theory of quantum gravity must incorporate
the Bekenestein-Hawking definition of black hole entropy \cite{1',2'} into its conceptual
framework. The black hole has a
thermodynamical entropy as
\begin{equation}\label{1'}
  S_{BH}=\frac{A}{4l_{p}^{2}}
\end{equation}
where $S_{BH}$ is the Bekenstein-Hawking entropy, $A$ is the area of the event horizon and $l_P=(\frac{\hbar G}{c^3})^{\frac{1}{2}}$ is the Planck length. One may be ask, what is the microscopic origin of this entropy? Although the various counting methods have pointed to the expected
semi-classical result, there is still a lack of recognition as to what degrees of freedom are
truly being counted.\\
Recently, motivated in great part by the recent works of Hawking, Perry and Strominger \cite{26a,6'}, it has appeared that a new way to approach the information paradox for black holes lies in a careful analysis of near horizon symmetries and the existence (in 4 dimensions) of an infinite-dimensional asymptotic symmetry group, the $bms_{4}$ group (For a good review and references, see \cite{7'}). Also, recently the authors of \cite{19} have obtained a very simple horizon entropy formula,
\begin{equation}\label{2'}
 S =2\pi (J_{0}^{+}+J_{0}^{-})
\end{equation}
for black hole solutions in $3D$ Einstein gravity, where $J_{0}^{\pm}$ are zero mode charges of $U(1)$ current algebra near horizon. In our previous paper \cite{3'} (see also \cite{4'}), we have studied the near horizon symmetry of spacelike warped AdS$_{3}$ black hole solutions of Generalized Minimal Massive Gravity (GMMG) \cite{140}. Similar to the near horizon symmetry algebra of the black flower solutions in Generalized minimal massive gravity \cite{5'}, the near horizon symmetry algebra of the warped black flower consists of two $U(1)$ current algebras, with different levels. We have shown that the formula \eqref{2'} exactly works for warped black flower solutions. So these investigations give us a non-trivial evidence for universality of this simple entropy formula in the context of 3D gravity. More recently an analog of the above entropy formula emerged in the near horizon description of non-extermal Kerr black holes in 4-dimensions \cite{18},
 \begin{equation}\label{3'}
 S =4\pi J_{0}^{+}J_{0}^{-}
\end{equation}
Now an interesting question is this :"if the simple entropy formula \eqref{3'} is again as universal as
its 3-dimensional pendant \eqref{2'}" \cite{18}. Therefore, in order to investigate universality of the entropy formula \eqref{3'} beyond pure Einstein gravity in 4-dimension, we consider the Einstein-Maxwell theory in the presence of cosmological constant. We show that this entropy formula give us the correct results for Kerr-Newman (A)dS black holes. We show that Kerr-Newman (A)dS black holes in 4-dimension Einstein-Maxwell exhibit an infinite-dimensional symmetry in their near horizon region.
\section{Conserved charges in Einstein-Maxwell Theory}\label{S.II}
First of all, we briefly review the approach of the covariant phase space method for obtaining conserved charges in Einstein-Maxwell Theory. To do this, we follow references \cite{1,2,3,4,5}. Suppose spacetime $(\mathcal{M},g)$ is globally hyperbolic and orientable. Suppose $\Phi$ is a collection of dynamical fields. The Lagrangian of Einstein-Maxwell theory is a functional of metric $g_{\mu \nu}$ and the gauge field $A_{\mu}$, then in this theory, we have $\Phi = \{ g_{\mu \nu} , A_{\mu} \}$. The Lagrangian of given theory is written as
\begin{equation}\label{1}
  L[\Phi]= \sqrt{-g} \mathcal{L}[\Phi],
\end{equation}
where
\begin{equation}\label{2}
  \mathcal{L}[\Phi]= R-2 \Lambda - 4 \pi F_{\mu \nu} F^{\mu \nu}.
\end{equation}
here $R$, $F_{\mu \nu} = \partial _{\mu} A_{\nu} - \partial _{\nu} A_{\mu} $ and $\Lambda$ are respectively the Ricci scalar, electromagnetic field strength and the cosmological constant. First order variation of the Lagrangian \eqref{1} is
\begin{equation}\label{3}
  \delta L [\Phi]=E_{\Phi}[\Phi] \delta \Phi + \partial_{\mu} \Theta ^{\mu}[\Phi , \delta \Phi],
\end{equation}
in which $E_{\Phi}$ have dual indices with $\Phi$ and sum on $\Phi$ is explicitly assumed. In the equation \eqref{3}, $\Theta ^{\mu}[\Phi , \delta \Phi]$ is the surface term and we refer to it as symplectic potential. Also, $E_{\Phi}=0$ give us the field equations. In Einstein-Maxwell theory, they are given as
\begin{equation}\label{4}
  E_{(g)}^{\mu \nu}= - \sqrt{-g} \left( G^{\mu \nu} + \Lambda g^{\mu \nu} - 8 \pi T^{\mu \nu} \right)=0,
\end{equation}
\begin{equation}\label{5}
  E_{(A)}^{\mu}=16 \pi \sqrt{-g} \nabla _{\nu} F^{\nu \mu}=0,
\end{equation}
\begin{equation}\label{6}
  \Theta^{\mu}[\Phi, \delta \Phi]= 2 \sqrt{-g} \left\{ \nabla ^{[\alpha} \left( g^{\mu] \beta} \delta g_{\alpha \beta} \right) -8 \pi F^{\mu \nu} \delta A_{\nu} \right\}.
\end{equation}
Equations \eqref{4} are known as Einstein's field equations, where $G^{\mu \nu}$ is the Einstein tensor and $T^{\mu \nu }$ is the electromagnetic energy-momentum tensor
\begin{equation}\label{7}
  T^{\mu \nu} = F^{\mu \alpha} F^{\nu}_{\hspace{1.7 mm} \alpha} - \frac{1}{4} g^{\mu \nu} F^{\alpha \beta} F_{\alpha \beta}.
\end{equation}
Also, equations \eqref{5} together with $ \nabla_{[\lambda}F_{\mu \nu]}=0 $ are Maxwell field equations in the curved spacetime.\\
Now consider two arbitrary variations $\delta_{1}$ and $\delta_{2}$. Suppose these two variations do not commute $\delta_{1} \delta_{2} \neq \delta_{2} \delta_{1}$. By varying Eq.\eqref{3}, we find second order variation of the Lagrangian
\begin{equation}\label{8}
  \delta_{1} \delta_{2} L [\Phi]=\delta_{1} E_{\Phi}[\Phi] \delta_{2} \Phi +  E_{\Phi}[\Phi] \delta_{1} \delta_{2} \Phi + \partial_{\mu} \delta_{1} \Theta ^{\mu}[\Phi , \delta _{2} \Phi].
\end{equation}
Similarly, one can write
\begin{equation}\label{9}
\delta_{2} \delta_{1} L [\Phi]=\delta_{2} E_{\Phi}[\Phi] \delta_{1} \Phi +  E_{\Phi}[\Phi] \delta_{2} \delta_{1} \Phi + \partial_{\mu} \delta_{2} \Theta ^{\mu}[\Phi , \delta _{1} \Phi].
\end{equation}
By subtracting Eq.\eqref{9} from Eq.\eqref{8}, we have
\begin{equation}\label{10}
  \delta_{[1,2]} L [\Phi]=\delta_{1} E_{\Phi}[\Phi] \delta_{2} \Phi - \delta_{2} E_{\Phi}[\Phi] \delta_{1} \Phi +  E_{\Phi}[\Phi] \delta_{[1,2]} \Phi + \partial_{\mu} \left( \delta_{1} \Theta ^{\mu}[\Phi , \delta _{2} \Phi] - \delta _{2} \Theta ^{\mu}[\Phi , \delta _{1} \Phi] \right),
\end{equation}
where $\delta_{[1,2]}=\delta_{1} \delta_{2} - \delta_{2} \delta_{1}$ is commutator of two variations $\delta_{1}$ and $\delta_{2}$. By using Eq.\eqref{3}, and replacing $\delta \rightarrow \delta_{[1,2]} $, we can write Eq.\eqref{10} as
\begin{equation}\label{11}
  \partial_{\mu} \omega_{\text{LW}}^{\mu}[\Phi; \delta_{1} \Phi , \delta_{2} \Phi]=- \frac{1}{16 \pi} \left( \delta_{1} E_{\Phi}[\Phi] \delta_{2} \Phi - \delta_{2} E_{\Phi}[\Phi] \delta_{1} \Phi \right),
\end{equation}
where
\begin{equation}\label{12}
   \omega_{\text{LW}}^{\mu}[\Phi; \delta_{1} \Phi , \delta_{2} \Phi]= \frac{1}{16 \pi} \left( \delta_{1} \Theta ^{\mu}[\Phi , \delta _{2} \Phi] - \delta _{2} \Theta ^{\mu}[\Phi , \delta _{1} \Phi] - \Theta ^{\mu}[\Phi , \delta _{[1,2]} \Phi] \right),
\end{equation}
is the Lee-Wald symplectic current. Since the symplectic potential is linear in $\delta \Phi$ then the terms containing $\delta _{1} \Phi$, $\delta _{2} \Phi$ and $\delta _{[1,2]} \Phi$ eliminate each other and $\omega_{\text{LW}}^{\mu}$ is a skew-symmetric bilinear in $\delta _{1} \Phi$ and $\delta _{2} \Phi$. The Lee-Wald symplectic current is conserved when equations of motion and linearized equations of motion are satisfied. In other words, if $\Phi$ is a solution of $E_{\Phi}=0$ and $\delta _{1} \Phi$ and $\delta _{2} \Phi$ are solutions of $\delta E_{\Phi}=0$, then the Lee-Wald symplectic current is conserved
\begin{equation}\label{13}
  \partial_{\mu} \omega_{\text{LW}}^{\mu}[\Phi; \delta_{1} \Phi , \delta_{2} \Phi] \simeq 0.
\end{equation}
The sign $\simeq$ indicates that the equality is held on-shell. We can define symplectic 2-form on solution space through the Lee-Wald symplectic current
\begin{equation}\label{14}
  \Omega_{\text{LW}} [\Phi; \delta_{1} \Phi , \delta_{2} \Phi]= \int_{\mathcal{C}} \omega_{\text{LW}}^{\mu}[\Phi; \delta_{1} \Phi , \delta_{2} \Phi] d^{3} x_{\mu},
\end{equation}
where $\mathcal{C}$ is a codimension-1 spacelike surface. Solution phase space can be constructed by factoring out the degeneracy subspace of configuration space (see Ref.\cite{1} for detailed discussion). Hence $\Omega_{\text{LW}}$ will be a symplectic form on solution phase space and it is closed, skew-symmetric and non-degenerate.\\
Suppose $\xi^{\mu}(x)$ and $\lambda(x)$ to be generators of diffeomorphism and $U(1)$ gauge transformation. We can introduce a combined transformation so that $\chi= (\xi , \lambda)$ is the generator of such transformations \cite{6}. The change in metric and $U(1)$ gauge field induced by an infinitesimal transformation generated by $\chi$ are given by
\begin{equation}\label{15}
  \delta _{\chi} g_{\mu \nu}= \mathcal{L}_{\xi} g_{\mu \nu},
\end{equation}
\begin{equation}\label{16}
  \delta _{\chi} A_{\mu}= \mathcal{L}_{\xi} A_{\mu} + \partial _{\mu} \lambda,
\end{equation}
respectively. Here, $\mathcal{L}_{\xi}$ denotes the Lie derivative along the vector field $\xi$. Also, the change in Lagrangian \eqref{1} induced by an infinitesimal transformation generated by $\chi$ is
\begin{equation}\label{17}
  \delta _{\chi} L[\Phi]= \mathcal{L}_{\xi} L[\Phi]= \partial_{\mu}\left( \xi^{\mu} L[\Phi] \right).
\end{equation}
 Since change in metric and $U(1)$ gauge field are linear in generator $\chi$ and change in the Lagrangian is a total derivative then $\chi$ generates a local symmetry on solution phase space \cite{1}. The generators of such local symmetry on solution phase space are conserved charges. The charge perturbation conjugate to $\chi$ is defined as
\begin{equation}\label{18}
  \delta Q_{\chi}= \Omega_{\text{LW}} [\Phi; \delta \Phi , \delta_{\chi} \Phi].
\end{equation}
The algebra among conserved charges is
\begin{equation}\label{19}
  \{ Q_{\chi_{1}}, Q_{\chi_{2}} \} = Q_{[\chi_{1},\chi_{2}]}+\tilde{\mathcal{C}}(\chi_{1},\chi_{2}),
\end{equation}
where $\tilde{\mathcal{C}}(\chi_{1},\chi_{2})$ is extension term and the Dirac bracket is defined as
\begin{equation}\label{20}
  \{ Q_{\chi_{1}}, Q_{\chi_{2}} \} = \delta_{\chi_{2}}Q_{\chi_{1}} .
\end{equation}
Now, we want to find explicit form of conserved charges in the Einstein-Maxwell theory. To this end, we assume that the variation in Eq.\eqref{3} is induced by an infinitesimal transformation generated by $\chi$
\begin{equation}\label{21}
  \delta _{\chi} L [\Phi] \simeq \partial_{\mu} \Theta ^{\mu}[\Phi , \delta _{\chi} \Phi],
\end{equation}
then we can define an on-shell Noether current
\begin{equation}\label{22}
  J_{\text{N}}^{\mu}[\Phi ; \chi] \simeq \Theta ^{\mu}[\Phi , \delta _{\chi} \Phi]- \xi^{\mu} L[\Phi],
\end{equation}
which is conserved on-shell, i.e. $\partial _{\mu}J_{\text{N}}^{\mu} \simeq 0 $. Thus there exists a second rank tensor density $K_{\text{N}}^{\mu \nu}[\Phi ; \chi]$ of weight $+1$ so that $ J_{\text{N}}^{\mu} \simeq \partial _{\nu} K_{\text{N}}^{\mu \nu}$. We refer to $K_{\text{N}}^{\mu \nu}$ as Noether potential and in the given theory it is given by
\begin{equation}\label{23}
  K_{\text{N}}^{\mu \nu}[\Phi ; \chi] \simeq - 2 \sqrt{-g} \left\{ \nabla ^{[\mu} \xi^{\nu]} +8 \pi F^{\mu \nu} \lambda \right\}.
\end{equation}
which can be obtained by substituting Eq.\eqref{1} and Eq.\eqref{6} into Eq.\eqref{22}. To find explicit form of the symplectic current, first, we take an arbitrary variation from Eq.\eqref{22}
\begin{equation}\label{24}
  \partial _{\nu} \delta K_{\text{N}}^{\mu \nu}[\Phi ; \chi] \simeq \delta \Theta ^{\mu}[\Phi , \delta _{\chi} \Phi]- \delta \left( \xi^{\mu} L[\Phi] \right).
\end{equation}
To have generality we assume that $\chi$ depends on the dynamical fields. On the other hand, second variation of Eq.\eqref{3}, induced by an infinitesimal transformation generated by $\chi$, is
\begin{equation}\label{25}
  \delta _{\chi} \delta L [\Phi] \simeq \partial_{\mu} \delta _{\chi} \Theta ^{\mu}[\Phi , \delta \Phi].
\end{equation}
 Since the commutator of an arbitrary variation and a variation induced by an infinitesimal transformation generated by $\chi$ is $ \delta \delta _{\chi}-\delta _{\chi} \delta = \delta _{\delta \chi}$ then the equation \eqref{25} can be written as
\begin{equation}\label{26}
  \delta \delta_{\chi} L [\Phi] - \delta _{\delta \chi} L [\Phi] \simeq \partial_{\mu} \delta _{\chi} \Theta ^{\mu}[\Phi , \delta \Phi].
\end{equation}
By substituting Eq.\eqref{21} into Eq.\eqref{26}, we find the explicit form of the symplectic current as
\begin{equation}\label{27}
    \omega_{\text{LW}}^{\mu} [\Phi; \delta \Phi , \delta_{\chi} \Phi] \simeq  \partial_{\nu} \mathcal{Q}_{\text{LW}}^{\mu \nu} [\Phi, \delta \Phi ;\chi ],
\end{equation}
with
\begin{equation}\label{28}
    \mathcal{Q}_{\text{LW}}^{\mu \nu} [\Phi, \delta \Phi ;\chi ] =\frac{1}{16 \pi} \left\{ \delta K_{\text{N}}^{\mu \nu}[\Phi ; \chi] - \delta K_{\text{N}}^{\mu \nu}[\Phi ; \delta \chi] + 2 \xi ^{[\mu} \Theta ^{\nu]}[\Phi , \delta \Phi ] \right\}.
\end{equation}
In the Einstein-Maxwell theory the explicit form of $\mathcal{Q}_{\text{LW}}^{\mu \nu}$ can be found by substituting equations \eqref{23} and \eqref{6} into the above equation
\begin{equation}\label{72}
\begin{split}
   \mathcal{Q}_{\text{LW}}^{\mu \nu} [\Phi, \delta \Phi ;\chi ] = & \frac{1}{8 \pi} \left[  h^{\lambda [ \mu} \nabla _{\lambda} \xi ^{\nu]} - \xi ^{\lambda} \nabla ^{[\mu} h^{\nu]}_{\lambda} - \frac{1}{2} h \nabla ^{[\mu} \xi ^{\nu]} + \xi ^{[\mu} \nabla _{\lambda} h^{\nu] \lambda} - \xi ^{[\mu} \nabla^{\nu]}h \right]\\
     & -2 \xi^{[\mu} F^{\nu] \alpha} \delta A_{\alpha} - \lambda \left( \delta F^{\mu \nu} + \frac{1}{2} h F^{\mu \nu} \right) ,
\end{split}
\end{equation}
where $h_{\mu \nu} = \delta g_{\mu \nu}$. The first line in the right hand side of Eq.\eqref{72} is the contribution from the gravity part and the second line is the contribution from the $U(1)$ gauge field part in the Lagrangian \eqref{1}. We can use Eq.\eqref{27} and Stokes' theorem to write conserved charge perturbation \eqref{18} as
\begin{equation}\label{29}
  \delta Q_{\chi}= \oint_{\mathcal{D}} \mathcal{Q}_{\text{LW}}^{\mu \nu} [\Phi, \delta \Phi ;\chi ] d^{2}x_{\mu \nu} ,
\end{equation}
where $\mathcal{D}$ denotes boundary of $\mathcal{C}$ and it is a spacelike codimension-2 surface. Usually it is thought that the linearization is just valid at spatial infinity. To overcome this problem, we take an integration from Eq.\eqref{29} over one-parameter path on the solution phase space. To this end, suppose that $\Phi (\mathcal{N})$ is a collection of fields which solve the equations of motion of the Einstein-Maxwell theory, where $\mathcal{N}$ is a free parameter in the solution phase space. Now, we replace $\mathcal{N}$ by $s\mathcal{N}$, where $0 \leq s \leq 1$ is just a parameter. By expanding $\Phi (s\mathcal{ N})$ in terms of $s$ we have $\Phi (s\mathcal{ N}) =  \Phi (0) + s \frac{\partial \Phi}{ \partial s} \big|_{s=0}  + \cdots$. By substituting $\Phi=\Phi (s\mathcal{ N})$ and $\delta \Phi =ds \frac{\partial \Phi}{ \partial s} \big|_{s=0}$ into Eq.\eqref{29}, we can define the conserved charge conjugate to $\chi$. Then we will have
\begin{equation}\label{30}
   Q_{\chi}= \int_{0}^{1} ds \oint_{\mathcal{D}} \mathcal{Q}_{\text{LW}}^{\mu \nu} [\Phi ;\chi | s ] d^{2}x_{\mu \nu} ,
\end{equation}
where integration over $s$ denotes integration over the one-parameter path on the solution phase space. In the equation \eqref{30}, $s=0$ is the value of the parameter corresponds to the background configuration. In this way, background contribution in the conserved charge is subtracted and then the conserved charge will be always finite. Therefore, this method is applicable to spacetimes with any backgrounds.
\section{Kerr-Newman (A)dS Black Hole}\label{S.III}
In this section we briefly review the Kerr-Newman (A)dS black hole geometry (see e.g. \cite{7}) and then we will find the near horizon behavior of this black hole in the Gaussian null coordinates.
\subsection{Geomerty}\label{S.III.A}
The Kerr-Newman (A)dS black hole is a solution for the Einstein-Maxwell theory. The metric corresponding to the given black hole in the Boyer-Lindquist coordinate system $(t, r, \theta, \hat{\phi})$ is
\begin{equation}\label{31}
  \begin{split}
     ds^{2}= & - \frac{\Delta _{r}-\Delta _{\theta} a^{2} \sin ^{2} \theta}{\Sigma} dt^{2}+ \frac{ 2 a \sin ^{2} \theta \left( \Delta _{r} - \Delta _{\theta} (r^2+ a^2) \right) }{ \Xi \Sigma } dt d \hat{\phi} + \frac{\Sigma}{\Delta _{r}} dr^2+ \frac{\Sigma}{\Delta _{\theta}} d\theta^2 \\
       & + \frac{\sin ^{2} \theta}{\Xi^{2} \Sigma} \left( \Delta _{\theta} (r^{2}+a^2)^2 - \Delta _{r} a^{2} \sin ^{2} \theta \right) d \hat{\phi}^2 ,
  \end{split}
\end{equation}
with
\begin{equation}\label{32}
  \Sigma = r^{2}+a^2 \cos ^{2} \theta , \hspace{1 cm} \Delta _{\theta} = 1+ \frac{\Lambda a^{2}}{3} \cos ^{2} \theta , \hspace{1 cm} \Xi = 1+ \frac{\Lambda a^{2}}{3},
\end{equation}
\begin{equation}\label{33}
  \Delta _{r}= \left( r^{2}+a^{2} \right) \left( 1-\frac{\Lambda}{3} r^{2} \right)-2 M r +\frac{Q^{2}}{4 \pi},
\end{equation}
where $M$, $a$ and $Q$ are constants and they parameterize the solution phase space. Also, the gauge field is given by
\begin{equation}\label{34}
  A=\frac{Q}{4 \pi} \frac{r}{\Sigma} \left(dt -\frac{a \sin ^{2} \theta}{\Xi} d \hat{\phi} \right).
\end{equation}
The mass, angular momentum and electric charge of the Kerr-Newman (A)dS black hole are related to $M$, $a$ and $Q$ as
\begin{equation}\label{35}
  \mathcal{M}=\frac{M}{\Xi^{2}}, \hspace{1 cm} \mathcal{J}=\frac{M a}{\Xi^{2}}, \hspace{1 cm} \mathcal{Q}=\frac{Q}{\Xi},
\end{equation}
respectively \cite{8}. Spacetime described by metric \eqref{31} is stationary and axially symmetric. Strictly speaking, it admits $\xi_{(t)}=\partial_{t}$ and $\xi_{(\hat{\phi})}=-\partial_{\hat{\phi}}$. Also, the theory considered in this paper is invariant under $U(1)$ gauge transformation. One can use covariant phase space method to find conserved charges conjugate to symmetry generators. According to the results of previous literature, mass, angular momentum and electric charge are conserved charges conjugate to symmetry generators $ \chi_{(t)} = (1,0,0,0,0)$, $ \chi_{(\hat{\phi})} = (0,0,0,-1,0)$ and $\chi_{(\lambda)} = (0,0,0,0,1)$, respectively. The Killing horizon-generating Killing vector field is $\zeta_{H}=\partial_{t} + \Omega_{H} \partial_{\hat{\phi}}$, where $\Omega_{H}$ denotes the horizon velocity. In gravity theories, conserved charges conjugate to the Killing vector $\zeta_{H}$ is proportional to product of surface gravity and black hole entropy. In Einstein-Maxwell theory this statement is no longer held and it has to be improved. Instead, conserved charge conjugate to the symmetry generator $\chi_{H}=(\zeta_{H}^{\mu}, \zeta_{H}^{\nu} A_{\nu})$ provides the desired result (for instance, see Refs. \cite{5,6,8}). The steps of calculating mass, angular momentum, electric charge and entropy of Kerr-Newman (A)dS Black Hole are exactly the same as what were done in \cite{5}. Therefore, we refrain from expressing the steps. The equation $ \Delta _{r}=0$ has at least two real roots $r_{+}$ and $r_{-}$ provided that the parameters $M$, $a$, $Q$, and $\Lambda$ are chosen suitably. The event horizon is located at $r=r_{+}$ (the largest real root of $\Delta _{r}$) and $r_{-}$ is the inner horizon radii. In this way we can rewrite $\Delta _{r}$ in the following form
\begin{equation}\label{36}
  \Delta _{r}= \left(r-r_{+}\right)\left(r-r_{-}\right) \left[ -\frac{\Lambda}{3} r^{2} -\frac{\Lambda}{3} (r_{+}+r_{-}) r+ \frac{1}{r_{+} r_{-}} \left( a^{2}+ \frac{Q^2}{4 \pi} \right) \right],
\end{equation}
where $M$ and $a$ are related to $r_{+}$ and $r_{-}$ as
\begin{equation}\label{37}
  M= \frac{(r_{+}+r_{-})}{2 \left( 1+\frac{\Lambda}{3} r_{+} r_{-}\right)} \left[\left( 1+\frac{\Lambda}{3} r_{+} r_{-}\right)^2 -\frac{\Lambda}{3} (r_{+}+r_{-})^2+\frac{\Lambda}{3} \frac{Q^2}{4 \pi} \right],
\end{equation}
\begin{equation}\label{38}
  a^{2}=\frac{1}{ \left( 1+\frac{\Lambda}{3} r_{+} r_{-}\right)} \left[r_{+}r_{-} \left( 1-\frac{\Lambda}{3} (r_{+}^{2}+r_{-}^{2}+ r_{+} r_{-}) \right)- \frac{Q^2}{4 \pi} \right].
\end{equation}
The above expressions will be reduced to the corresponding expressions for Kerr-Newman black hole when we set $\Lambda=0$.
\subsection{Near horizon behaviour}\label{S.III.B}
In order to find near horizon geometry of the Kerr-Newman (A)dS black hole first we write the metric \eqref{31} in the advanced Eddington-Finkelstein coordinates $(v,r,\theta, \tilde{\phi})$. To this end, we transform coordinates as
\begin{equation}\label{39}
  dv=dt+ \frac{(r^{2}+a^{2})}{\Delta _{r}} dr , \hspace{1 cm} d \tilde{\phi}= d \hat{\phi}+ \frac{a \Xi}{\Delta _{r}} dr.
\end{equation}
and we find that
\begin{equation}\label{40}
  \begin{split}
     ds^{2}= & - \frac{\Delta _{r}-\Delta _{\theta} a^{2} \sin ^{2} \theta}{\Sigma} dv^{2} +2 dv dr+ \frac{ 2 a \sin ^{2} \theta \left( \Delta _{r} - \Delta _{\theta} (r^2+ a^2) \right) }{ \Xi \Sigma } dv d \tilde{\phi} + \frac{\Sigma}{\Delta _{\theta}} d\theta^2 \\
       & - \frac{2 a \sin ^{2} \theta}{\Xi} dr d \tilde{\phi}+ \frac{\sin ^{2} \theta}{\Xi^{2} \Sigma} \left( \Delta _{\theta} (r^{2}+a^2)^2 - \Delta _{r} a^{2} \sin ^{2} \theta \right) d \tilde{\phi}^2 .
  \end{split}
\end{equation}
 Since the horizon velocity is given by $\Omega_{H}= \frac{a \Xi}{r_{+}^{2}+a^2}$ then it is natural that we perform another coordinates transformation as $\tilde{\phi}= \phi + \Omega_{H} v$. Here index $H$ refers to the Horizon. In this way, we have
\begin{equation}\label{41}
  \begin{split}
     ds^{2}= & - \frac{\Sigma_{H}^{2} \Delta _{r}- (r^{2}-r_{+}^2)^2 \Delta _{\theta} a^{2} \sin ^{2} \theta }{\Sigma (r_{+}^{2}+a^2)^2} dv^{2} + \frac{ 2 a \sin ^{2} \theta \left( \Sigma_{H} \Delta _{r} + \Delta _{\theta} (r^2+ a^2) (r^{2}-r_{+}^2)\right) }{ \Xi \Sigma (r_{+}^2+ a^2)} dv d \phi  \\
       & + \frac{2 \Sigma_{H}}{(r_{+}^{2}+a^2)} dv dr - \frac{2 a \sin ^{2} \theta}{\Xi} dr d \phi + \frac{\Sigma}{\Delta _{\theta}} d\theta^2+ \frac{\sin ^{2} \theta}{\Xi^{2} \Sigma} \left( \Delta _{\theta} (r^{2}+a^2)^2 - \Delta _{r} a^{2} \sin ^{2} \theta \right) d \phi^2 .
  \end{split}
\end{equation}
In this coordinate system the $U(1)$ gauge field can be written as
\begin{equation}\label{42}
  A=\frac{Q}{4 \pi} \frac{r}{\Sigma} \left(\frac{\Sigma_{H}}{(r_{+}^{2}+a^2)} dv -\frac{a \sin ^{2} \theta}{\Xi} d \phi \right),
\end{equation}
and we have $g_{vv} =g_{v \phi} =0$ on the event horizon. Now we write near horizon fall-off conditions for Kerr-Newman (A)dS black hole in the Gaussian null coordinate system. To do this, we follow the method proposed in Appendix A of the paper \cite{9}. Therefore, we rewrite the metric relative to the correct set of geodesics. A suitable pair of cross-normalized null normals is
\begin{equation}\label{43}
  l= \partial_{v}, \hspace{1 cm} \text{and} \hspace{1 cm} n= \frac{a^{2} \sin ^{2} \theta }{2 \Delta _{\theta} \Sigma _{H}} \partial _{v}+ \frac{(r_{+}^{2} +a^{2})}{\Sigma _{H}} \partial _{r}+ \frac{a \Xi}{\Delta _{\theta} (r_{+}^{2} +a^{2})} \partial_{\phi}.
\end{equation}
These vectors are defined on horizon and we have $l \cdot l |_{H}=n \cdot n |_{H}=0$ and $l \cdot n =1$. Now we consider a family of null geodesics that crosses $H$. The vector field tangent to them is $n$ and they are labeled by $(v, \theta, \phi)$. Suppose $\rho$ is an affine parameter which parameterize the given geodesics so that $\rho=0$ on $H$. The geodesics can be constructed up to third order in $\rho$:
\begin{equation}\label{44}
  X_{(v,\theta, \phi)}^{\mu}(\rho)= X^{\mu}\big|_{\rho =0} + \rho \frac{d X^{\mu}}{d \rho} \bigg|_{\rho=0}+ \frac{1}{2} \rho^{2} \frac{d^{2} X^{\mu}}{d \rho^{2}} \bigg|_{\rho=0}+ \mathcal{O}(\rho^{3}),
\end{equation}
where $X^{\mu}\big|_{\rho =0} = (v,r_{+},\theta, \phi)$ and $\frac{d X^{\mu}}{d \rho} \big|_{\rho=0}=n^{\mu}$. Also, by using geodesic equation $n^{\nu} \nabla _{\nu} n^{\mu} |_{H}=0$, one can find the second order derivative of $X^{\mu}$ with respect to $\rho$ at horizon
\begin{equation}\label{45}
  \frac{d^{2} X^{\mu}}{d \rho^{2}} \bigg|_{\rho=0} = - \Gamma^{\mu}_{\alpha \beta} n^{\alpha} n^{\beta} \bigg|_{\rho=0}.
\end{equation}
The equation \eqref{44} defines a transformation from $(v,r,\theta, \phi)$ to $(v, \rho, \theta , \phi)$ and then we can calculate the first order expansion of the metric $g_{\mu \nu} = g_{\mu \nu}^{(0)}+ \rho g_{\mu \nu}^{(1)}+ \mathcal{O}(\rho^{2})$, where
\begin{equation}\label{46}
  g_{v \rho}^{(0)}= 1, \hspace{1 cm} g_{\theta \theta}^{(0)}= \frac{\Sigma _{H}}{\Delta _{\theta}} , \hspace{1 cm} g_{\phi \phi}^{(0)}= \frac{\Delta _{\theta} (r_{+}^{2}+a^{2})^{2} \sin ^{2} \theta}{\Xi ^{2} \Sigma _{H}},
\end{equation}
\begin{equation}\label{47}
  g_{vv}^{(1)}=-2 \kappa , \hspace{0.8 cm} g_{v \theta}^{(1)}= \frac{2 a^{2} \sin \theta \cos \theta}{\Sigma _{H}} , \hspace{0.8 cm} g_{v \phi}^{(1)}= \frac{a \sin ^{2} \theta}{ \Xi \Sigma_{H}^{2} } \left[ \Sigma _{H} \Delta_{r}^{\prime}(r_{+})+2 r_{+} \Delta_{\theta} (r_{+}^{2}+a^{2}) \right],
\end{equation}
\begin{equation}\label{48}
  \begin{split}
      & g_{\theta \theta}^{(1)}= \frac{2 r_{+} (r_{+}^{2}+a^{2})}{\Delta _{\theta} \Sigma_{H}}, \hspace{1 cm} g_{\phi \phi}^{(1)}= \frac{2 r_{+} (r_{+}^{2}+a^{2})^{2} \Delta_{\theta} \sin ^{2} \theta}{\Xi^{2} \Sigma_{H}^{3}} \left( 2 \Sigma_{H}-(r_{+}^{2}+a^{2})\right), \hspace{1 cm} \\
       & g_{\theta \phi}^{(1)}= - \frac{2 a^{3} (r_{+}^{2}+a^{2}) \sin^{3} \theta \cos \theta}{\Xi \Sigma _{H}^{2} \Delta_{\theta}} \left(1-\frac{\Lambda}{3} r_{+}^{2} \right),
  \end{split}
\end{equation}
here the prime denotes derivative with respect to radial coordinate. Also, $\kappa$ is surface gravity of the Kerr-Newman (A)dS black hole
\begin{equation}\label{49}
  \kappa = \frac{\Delta_{r}^{\prime}(r_{+})}{2 (r_{+}^{2}+a^{2})}.
\end{equation}
In the new coordinate system, the gauge field can be expanded as
\begin{equation}\label{50}
\begin{split}
     & A_{v}= \frac{Q}{4 \pi} \frac{r_{+}}{(r_{+}^{2}+a^{2})} - \frac{Q}{4 \pi} \frac{(r_{+}^{2} - a^{2} \cos^{2} \theta)}{\Sigma_{H}^{2}} \rho +\mathcal{O}(\rho^{2}), \hspace{1 cm} A_{\phi}= - \frac{Q}{4 \pi} \frac{a r_{+} \sin^{2} \theta}{\Xi\Sigma_{H}}+\mathcal{O}(\rho),\\
     &  A_{\rho}=\mathcal{O}(\rho^{2}), \hspace{1 cm} A_{\theta}= \mathcal{O}(\rho) .
\end{split}
\end{equation}
We have avoided writing the first order terms in $A_{\phi}$ and $A_{\theta}$ because we do not need them. From Eq.\eqref{46}, the full 2-metric on horizon is
\begin{equation}\label{51}
  d \sigma ^{2}= \frac{\Sigma _{H}}{\Delta _{\theta}} d \theta^{2}+ \frac{\Delta _{\theta} (r_{+}^{2}+a^{2})^{2} \sin ^{2} \theta}{\Xi ^{2} \Sigma _{H}} d \phi^{2}.
\end{equation}
This metric is conformally related to Riemann sphere. To show this relation, we introduce a field-dependent change of coordinates
\begin{equation}\label{52}
  z= e^{i \phi} \mu (\theta), \hspace{1 cm} \bar{z}= e^{-i \phi} \mu (\theta),
\end{equation}
where $\mu (\theta)$ is a real function of $\theta$ and $\bar{z}$ is complex conjugate to $z$. Depending on the sign of cosmological constant, $\mu (\theta)$ will be different. The explicit form of $\mu (\theta)$ can be written as
\begin{equation}\label{53}
  \mu (\theta)= e^{-\frac{a^{2}}{r_{+}^{2}+a^{2}} \mathcal{W}(\theta)} \cot (\theta /2),
\end{equation}
where $\mathcal{W}(\theta)$ is a function of $\theta$ and its explicit form depends on the sign of cosmological constant
\begin{equation}\label{54}
  \mathcal{W}(\theta) = \left\{
     \begin{array}{lr}
       \cos \theta & \text{for} \hspace{1 cm} \Lambda =0 \\
       \frac{l}{a} \left( 1- \frac{r_{+}^{2}}{l^{2}}\right) \tan^{-1} \left( \frac{a}{l} \cos \theta \right) & \text{for} \hspace{1 cm} \Lambda =\frac{3}{l^{2}} > 0 \\
       \frac{l}{a} \left( 1+ \frac{r_{+}^{2}}{l^{2}}\right) \tanh^{-1} \left( \frac{a}{l} \cos \theta \right) & \text{for} \hspace{1 cm} \Lambda =-\frac{3}{l^{2}} < 0 \\
     \end{array}
   \right.
\end{equation}
The both $\Lambda > 0$ and $\Lambda < 0 $ cases will tend to $\Lambda =0$ case when $l \rightarrow \infty$. Now, we can write the metric of the horizon in the conformal form
\begin{equation}\label{55}
\begin{split}
    d \sigma^{2} =& \Omega \gamma_{AB} dx^{A} dx^{B} \\
     =& \Omega \frac{4 dz d\bar{z}}{\left( 1+z \bar{z}\right)^{2}},
\end{split}
\end{equation}
with
\begin{equation}\label{56}
  \Omega= \frac{\Delta_{\theta} (r_{+}^{2}+a^{2})^{2}}{\Xi ^{2} \Sigma _{H}} \left( e^{-\frac{a^{2}}{r_{+}^{2}+a^{2}} \mathcal{W}(\theta)} \cos^{2} (\theta / 2)+ e^{+\frac{a^{2}}{r_{+}^{2}+a^{2}} \mathcal{W}(\theta)} \sin^{2} (\theta / 2) \right)^{2}.
\end{equation}
The conformal factor $\Omega$ is a function of $z$ and $\bar{z}$. Hence, the metric of the horizon is locally, conformally equivalent to the two-sphere.
\section{Near horizon fall-off conditions}\label{S.IV}
In the previous section, we wrote the near horizon metric in the Gaussian null coordinate system. Therefore, following \cite{10}, we can consider near horizon fall-off conditions for the Kerr-Newman (A)dS black hole to be
\begin{equation}\label{57}
  ds^{2}= -2 \kappa \rho dv^{2}+2 dv d\rho + 2 \rho \theta_{A} dv dx^{A} +(\Omega_{AB}+ \rho \lambda_{AB}) dx^{A} dx^{B} + \mathcal{O}(\rho^{2}),
\end{equation}
where $v$ is the advanced time coordinate such that a null surface is defined by $g^{\alpha \beta} \partial_{\alpha}v \partial_{\beta}v =0$ and the vector tangent to this surface is given by $k^{\mu} =g^{\mu \nu} \partial_{\nu}v$ which defines a ray. Also, $\rho$ is the affine parameter of the
generator $k^{\mu}$. Suppose $\kappa$, $\theta_{A}$, $\Omega_{AB}$ and $\lambda_{AB}$ are functions of $x^{A}$, where two coordinates $x^{A}$ are chosen constant along each ray. Also, one can introduce following near horizon fall-off conditions for the $U(1)$ gauge field
\begin{equation}\label{58}
  A= \left( \varphi_{v}+ \rho \psi_{v} \right)dv+\left( \varphi_{A}+ \rho \psi_{A} \right)dx^{A}+\mathcal{O}(\rho^{2}),
\end{equation}
where we set $A_{\rho}=0$ as a gauge condition and $\varphi_{v}$, $\psi_{v}$, $\varphi_{A}$ and $\psi_{A}$ are functions of $x^{A}$.\\
By substituting fall-off conditions \eqref{57} and \eqref{58} into the field equations, we can find additional restrictions. The $(v,A) $ components of Einstein's field equations at zeroth order restrict $\kappa$ to be a constant, i.e. $\kappa$ is independent of $x^{A}$. Also, the $(v,v) $ component of Einstein's field equations at zeroth order yields
\begin{equation}\label{59}
  \Omega^{AB} \partial_{A} \varphi_{v} \partial_{B}\varphi_{v}=0,
\end{equation}
where $\Omega^{AB}$ is the inverse of $\Omega_{AB}$ (we explicitly assume that $\Omega_{AB}$ is invertible). Since the metric of horizon $\Omega_{AB}$ is a Riemannian (not Lorentzian) one, then $\varphi_{v}$ has to be a constant. The other components of the equations of motion relate first order terms to zeroth order ones in metric and gauge field expansions and we do not need them later.
\section{Near horizon symmetries}\label{S.V}
The change in metric and $U(1)$ gauge field induced by an infinitesimal transformation generated by $\chi$ are given by \eqref{15} and \eqref{16}. Now, we want to find the residual symmetries such that they respect to fall-off conditions \eqref{57} and \eqref{58}. We find that symmetry generator $\chi$ with following components
\begin{equation}\label{60}
\begin{split}
     & \xi^{v}= T, \hspace{1 cm} \xi^{\rho}= \frac{1}{2} \rho^{2} \Omega^{AB} \theta_{A} \partial _{B} T+ \mathcal{O}(\rho^{3}), \\
     & \xi^{A}= Y^{A} - \rho \Omega^{A B} \partial _{B} T+\frac{1}{2} \rho^{2} \Omega^{AC} \Omega^{BD} \lambda_{C D} \partial _{B} T+ \mathcal{O}(\rho^{3}),
\end{split}
\end{equation}
\begin{equation}\label{61}
  \lambda= \hat{\lambda}+ \rho \Omega^{AB} \varphi_{A} \partial _{B} T - \frac{1}{2} \rho^{2} \left( \Omega^{AC} \Omega^{BD} \lambda_{C D} \varphi_{A} \partial _{B} T- \Omega^{AB}\psi_{A} \partial _{B} T\right) + \mathcal{O}(\rho^{3}),
\end{equation}
preserves the given near horizon fall-off conditions. Here $T$, $Y^{A}$ and $\hat{\lambda}$ are arbitrary functions of $x^{A}$. In order to obtain the asymptotic symmetry generator $\chi$, we assumed that the leading terms does not depend on the dynamical fields. Under such an assumption the boundary conditions will be "state independent", which means that the form of the asymptotic symmetry generators are not considered to depend explicitly of the charges \cite{12}. The change in dynamical fields under the action of symmetry generator $\chi$ can be read as
\begin{equation}\label{62}
  \begin{split}
       & \delta_{\chi} \theta_{A}= \mathcal{L}_{Y} \theta_{A}-2 \kappa \partial_{A} T, \hspace{1 cm}  \delta_{\chi} \Omega_{AB} =\mathcal{L}_{Y}\Omega_{AB} , \\
       & \delta_{\chi} \lambda_{AB} = \mathcal{L}_{Y} \lambda_{AB} + \theta_{A} \partial _{B} T+ \theta_{B} \partial _{A} T - 2 \bar{\nabla}_{A}\bar{\nabla}_{B} T,
  \end{split}
\end{equation}
\begin{equation}\label{63}
  \begin{split}
       & \delta_{\chi} \psi_{v}= \mathcal{L}_{Y} \psi_{v}, \hspace{1 cm}  \delta_{\chi} \varphi_{A} =\mathcal{L}_{Y}\varphi_{A}+ \varphi_{v} \partial_{A} T+ \partial_{A} \hat{\lambda}, \\
       & \delta_{\chi} \psi_{A} = \mathcal{L}_{Y} \psi_{A} + \psi_{v} \partial_{A} T + \Omega^{BC} \left( \partial_{A} \varphi_{B}-\partial_{B} \varphi_{A} \right) \partial_{C}T,
  \end{split}
\end{equation}
where $\mathcal{L}_{Y}$ denotes the Lie derivative along $Y^{A}$ and $\bar{\nabla}_{A}$ is the covariant derivative with respect to connection $\bar{\Gamma}^{A}_{BC}$ compatible with the metric of the horizon $\Omega_{AB}$. It is worth mentioning that because $\kappa$ and $\varphi_{v}$ are not dynamical then they will remain unchanged under the action of the symmetry generator $\chi$, i.e. $\delta_{\chi}\kappa =0$ and $\delta_{\chi}\varphi_{v}=0$.\\
The asymptotic Killing vectors \eqref{60} are functions of the dynamical fields. To take it into account we introduce a modified version of Lie brackets \cite{11}
\begin{equation}\label{64}
  \left[ \xi_{1} , \xi_{2} \right] =  \mathcal{L} _{\xi_{1}} \xi_{2} - \delta ^{(g)}_{\xi _{1}} \xi_{2} + \delta ^{(g)}_{\xi _{2}} \xi_{1},
\end{equation}
where $\xi_{1}= \xi(T_{1}, Y^{A}_{1})$ and $\xi_{2}= \xi(T_{2}, Y^{A}_{2})$ and $\delta ^{(g)}_{\xi _{1}} \xi_{2}$ denotes the change induced in $\xi_{2}$ due to the variation of metric $\delta _{_{\xi _{1}}} g_{\mu\nu} = \mathcal{L} _{\xi_{1}} g_{\mu\nu}$. Therefore one finds that
\begin{equation}\label{65}
  \left[ \xi_{1} , \xi_{2} \right] = \xi_{12},
\end{equation}
with $\xi_{12}= \xi(T_{12}, Y^{A}_{12})$, where
\begin{equation}\label{66}
  T_{12}= Y_{1}^{A} \partial_{A}T_{2} - Y_{2}^{A} \partial_{A}T_{1}, \hspace{1 cm} Y_{12}^{A}= Y_{1}^{B} \partial_{B}Y_{2}^{A}- Y_{2}^{B} \partial_{B}Y_{1}^{A}.
\end{equation}
Thus, the algebra of asymptotic Killing vectors is closed. In addition to $T$ and $Y^{A}$, the symmetry generator $\chi=\chi (T, Y^{A}, \hat{\lambda})$ contains another degree of freedom, $\hat{\lambda}$. Here, $\hat{\lambda}$ is an arbitrary function of $x^{A}$ and generates $U(1)$ symmetry. Hence, we need to introduce two other commutators
\begin{equation}\label{67}
  [\chi(0,0,0,\hat{\lambda}_{1}), \chi(0,0,0,\hat{\lambda}_{2})]= 0,
\end{equation}
\begin{equation}\label{68}
  [\chi(0,0,0,\hat{\lambda}_{1}), \chi(0,Y_{2}^{A},0)] =-[\chi(0,Y_{2}^{A},0),\chi(0,0,0,\hat{\lambda}_{1})]=\chi(0,0,0, -\mathcal{L}_{Y_{2}} \hat{\lambda}_{1}),
\end{equation}
in addition to Eq.\eqref{65}. The equation \eqref{67} comes from the fact that $U(1)$ is an Abelian group and we will justify Eq.\eqref{68} when we consider the algebra among conserved charges.\\
The induced metric on the horizon $\Omega_{AB}$ is conformally related to the Riemann sphere and the Laurent expansion on the Riemann sphere is allowed. Since the general solution of the conformal Killing equations is $Y= Y^{z}(z) \partial_{z}+Y^{\bar{z}}(\bar{z}) \partial_{\bar{z}}$ and $T= T(z,\bar{z})$ and $\hat{\lambda}= \hat{\lambda}(z,\bar{z})$ are arbitrary functions of $z$ and $\bar{z}$, we can define modes as
\begin{equation}\label{69}
  \begin{split}
       & T_{(m,n)}=\chi (z^{m}\bar{z}^{n},0,0,0), \hspace{1 cm} Y_{m}=\chi (0,-z^{m+1},0,0), \hspace{1 cm} \bar{Y}_{m}=\chi (0,0,-\bar{z}^{m+1},0), \\
       & \hat{\lambda}_{(m,n)}=\chi (0,0,0,z^{m}\bar{z}^{n}),
  \end{split}
\end{equation}
where $m,n \in \mathbb{Z}$. By using equations \eqref{65}, \eqref{67} and \eqref{68}, we find the algebra among these modes
\begin{equation}\label{70}
  \begin{split}
       & [Y_{m}, Y_{n}]= (m-n) Y_{m+n}, \hspace{1 cm} [\bar{Y}_{m}, \bar{Y}_{n}]= (m-n) \bar{Y}_{m+n}, \hspace{1 cm} [Y_{m}, \bar{Y}_{n}]=0, \\
       & [T_{(m,n)}, T_{(k,l)}]=0, \hspace{1 cm} [Y_{k}, T_{(m,n)} ]= -m T_{(m+k,n)}, \hspace{1 cm} [\bar{Y}_{k}, T_{(m,n)} ]= -n T_{(m,n+k)},
  \end{split}
\end{equation}
\begin{equation}\label{71}
  \begin{split}
       & [\hat{\lambda}_{(m,n)}, \hat{\lambda}_{(k,l)}]=0, \hspace{1 cm} [Y_{k}, \hat{\lambda}_{(m,n)} ]= -m \hat{\lambda}_{(m+k,n)}, \hspace{1 cm} [\bar{Y}_{k}, \hat{\lambda}_{(m,n)} ]= -n \hat{\lambda}_{(m,n+k)},\\
       & [\hat{\lambda}_{(m,n)}, T_{(k,l)}]=0,
  \end{split}
\end{equation}
This algebra contains a set of supertranslations current $T_{(m,n)}$ and two sets of Witt algebra currents, given by $Y_{m}$ and $\bar{Y}_{m}$. It also contains a set of multiple-charges current $\hat{\lambda}_{(m,n)}$. Two sets of Witt currents are in semi-direct sum with the supertranslations and multiple-charges current. The subalgebra \eqref{70} is known as $\mathfrak{bms}_{4}^{H}$ \cite{12,16} and it differs from Bondi-Metzner-Sachs algebra $\mathfrak{bms}_{4}$ \cite{13,14,15}(the structure constants are different).
\section{Charges and Soft Hairs}\label{S.VI}
Now we are ready to find conserved charge conjugate to the asymptotic symmetry generator $\chi$ with components \eqref{60} and \eqref{61}. To this end, we take codimension-two surface $\mathcal{D}$ in Eq.\eqref{29} to be the horizon
\begin{equation}\label{73}
\begin{split}
   \delta Q_{\chi}= & \oint_{H} \mathcal{Q}_{\text{LW}}^{\mu \nu} [\Phi, \delta \Phi ;\chi ] d^{2}x_{\mu \nu} , \\
    = & \int d^{2} x \sqrt{\det \Omega} \mathcal{Q}_{\text{LW}}^{v \rho}\big|_{\rho=0}.
\end{split}
\end{equation}
By substituting the boundary conditions and components of the asymptotic symmetry generators into Eq.\eqref{73}, we have
\begin{equation}\label{74}
  Q_{\chi}= \frac{1}{8 \pi} \int d^{2} x \sqrt{\det \Omega} \left( \kappa T - \frac{1}{2} Y^{A} \theta_{A} - 8 \pi \hat{\lambda} \psi_{v}\right),
\end{equation}
where an integral over one-parameter path on solution phase space was taken. As we mentioned earlier, one can use equations \eqref{19} and \eqref{20} to find the algebra among the conserved charges. After performing some calculations, we find that
\begin{equation}\label{75}
  \left\{ Q_{\chi_{1}}, Q_{\chi_{2}} \right\}= Q_{[\chi_{1}, \chi_{2}]},
\end{equation}
where equations \eqref{62} and \eqref{63} were used also $[\chi_{1}, \chi_{2}]$ is given by equations \eqref{65}-\eqref{68}. In this case, by comparing Eq.\eqref{19} and \eqref{75}, we see that the central extension term does not appear. Since the algebra among the conserved charges is isomorphic to algebra among symmetry generators and the commutation relation \eqref{68} is appeared in the right hand side of Eq.\eqref{75} then it seems reasonable to consider such a commutation relation. By substituting Eq.\eqref{69} into Eq.\eqref{74}, supertranslation, superrotation and multiple-charge modes can be obtained as
\begin{equation}\label{76}
  \mathcal{T}_{(m,n)}= \frac{\kappa}{8 \pi} \int dz d \bar{z} \Omega \sqrt{\gamma} z^{m} \bar{z}^{n},
\end{equation}
\begin{equation}\label{77}
    \mathcal{Y}_{m}= \frac{1}{16 \pi} \int dz d \bar{z} \Omega \sqrt{\gamma} z^{m+1} \theta_{z},
\end{equation}
\begin{equation}\label{78}
    \bar{\mathcal{Y}}_{m}=  \frac{1}{16 \pi} \int dz d \bar{z} \Omega \sqrt{\gamma} \bar{z}^{m+1} \theta_{\bar{z}},
\end{equation}
\begin{equation}\label{79}
  \mathcal{Q}_{(m,n)}= - \int dz d \bar{z} \Omega \sqrt{\gamma} z^{m} \bar{z}^{n} \psi_{v},
\end{equation}
respectively, where $\gamma = \det (\gamma_{AB})$. Also, the equation \eqref{75} gives us the algebra among these modes
\begin{equation}\label{80}
  \begin{split}
       & \{ \mathcal{Y}_{m}, \mathcal{Y}_{n}\}= (m-n) \mathcal{Y}_{m+n}, \hspace{1 cm} \{ \bar{\mathcal{Y}}_{m}, \bar{\mathcal{Y}}_{n}\}= (m-n) \bar{\mathcal{Y}}_{m+n}, \hspace{1 cm} \{ \mathcal{Y}_{m}, \bar{\mathcal{Y}}_{n}\}=0, \\
       & \{ \mathcal{T}_{(m,n)}, \mathcal{T}_{(k,l)} \}=0, \hspace{1 cm} \{ \mathcal{Y}_{k}, \mathcal{T}_{(m,n)} \} = -m \mathcal{T}_{(m+k,n)}, \hspace{1 cm} \{ \bar{\mathcal{Y}}_{k}, \mathcal{T}_{(m,n)} \}= -n \mathcal{T}_{(m,n+k)},
  \end{split}
\end{equation}
\begin{equation}\label{81}
  \begin{split}
       & \{\mathcal{Q}_{(m,n)}, \mathcal{Q}_{(k,l)}\}=0, \hspace{1 cm} \{\mathcal{Y}_{k}, \mathcal{Q}_{(m,n)} \}= -m \mathcal{Q}_{(m+k,n)}, \hspace{1 cm} \{\bar{\mathcal{Y}}_{k}, \mathcal{Q}_{(m,n)} \}= -n \mathcal{Q}_{(m,n+k)},\\
       & \{\mathcal{Q}_{(m,n)}, \mathcal{T}_{(k,l)}\}=0.
  \end{split}
\end{equation}
Now we apply the above considerations on the Kerr-Newman (A)dS black hole. Consequently, we can obtain charge zero-modes explicitly and then interpret them. Since $(z,\bar{z})$ coordinates are related to $(\theta , \phi)$ coordinates through the equation \eqref{52} and
\begin{equation}\label{82}
  dz d \bar{z} \Omega \sqrt{\gamma}=\frac{(r_{+}^{2}+a^{2})}{\Xi} \sin \theta d \theta d \phi,
\end{equation}
then we can easily read supertranslation charge modes as
\begin{equation}\label{83}
  \mathcal{T}_{(m,n)}= \frac{\kappa (r_{+}^{2}+a^{2})}{4 \Xi} I(m) \delta_{m,n},
\end{equation}
where
\begin{equation}\label{84}
  I(m)= \int_{0}^{\pi} \mu (\theta)^{2m} \sin \theta d \theta .
\end{equation}
Hence, the supertranslation double-zero-mode charge $\mathcal{T}_{(0,0)}$ is
\begin{equation}\label{85}
  \mathcal{T}_{(0,0)}= \frac{\kappa (r_{+}^{2}+a^{2})}{2 \Xi},
\end{equation}
where $I(0)=2$ was used. The entropy of the Kerr-Newman (A)dS black hole is given by (see e.g. \cite{8})
\begin{equation}\label{86}
  S=\frac{ \pi (r_{+}^{2}+a^{2})}{ \Xi}
\end{equation}
then the supertranslation double-zero-mode charge $\mathcal{T}_{(0,0)}$ is equal to the the Kerr-Newman (A)dS black hole entropy multiplied by Hawking temperature $T_{H}= \kappa/2 \pi$, as expected \cite{12,16,17}. Similarly, multiple-charge modes are
\begin{equation}\label{87}
  \mathcal{Q}_{(m,n)}=\frac{Q (r_{+}^{2}+a^{2})}{2 \Xi} W(m) \delta_{m,n},
\end{equation}
with
\begin{equation}\label{88}
  W(m)= \int_{0}^{\pi} \frac{(r_{+}^{2}-a^{2} \cos^{2} \theta)}{(r_{+}^{2}+a^{2} \cos^{2} \theta)^{2}} \mu (\theta)^{2m} \sin \theta d \theta .
\end{equation}
where $\psi_{v}= A^{(1)}_{v}$ was used. One can show that the multiple-charge double-zero-mode gives the Kerr-Newman (A)dS black hole electric charge
\begin{equation}\label{89}
  \mathcal{Q}_{(0,0)}=\frac{Q}{\Xi}.
\end{equation}
Now we calculate superrotation charges. $\theta_{z}$ and $\theta_{\bar{z}}$ are related to $\theta_{\theta}$ and $\theta_{\phi}$ as
\begin{equation}\label{90}
     z \theta_{z}= \frac{1}{2} \left( \frac{\mu}{\partial_{\theta} \mu } \theta_{\theta} -i \theta_{\phi} \right), \hspace{1 cm}  \bar{z} \theta_{\bar{z}}= \frac{1}{2} \left( \frac{\mu}{\partial_{\theta} \mu } \theta_{\theta} +i \theta_{\phi} \right).
\end{equation}
These relations are deduced from the fact that $\theta_{z}dz + \theta_{\bar{z}}d\bar{z}= \theta_{\theta}d\theta + \theta_{\phi}d\phi$. Also, we have $\theta_{\theta}= g_{v \theta}^{(1)}$ and $\theta_{\phi}= g_{v \phi}^{(1)}$. Therefore, the explicit form of the charges associated to superrotations are
\begin{equation}\label{91}
  \mathcal{Y}_{m}=-i \frac{M a}{2 \Xi ^{2}} \delta_{m,0}+ i \frac{Q^{2}}{4 \pi} \frac{(r_{+}^{2}+a^{2})^{2} \tan^{-1} (\frac{a}{r_{+}}) - r_{+} a (r_{+}^{2}-a^{2})}{8 \Xi^{2} r_{+}^{2} a^{2}} \delta_{m,0},
\end{equation}
\begin{equation}\label{92}
  \bar{\mathcal{Y}}_{m}=+i \frac{M a}{2 \Xi ^{2}} \delta_{m,0}- i \frac{Q^{2}}{4 \pi} \frac{(r_{+}^{2}+a^{2})^{2} \tan^{-1} (\frac{a}{r_{+}}) - r_{+} a (r_{+}^{2}-a^{2})}{8 \Xi^{2} r_{+}^{2} a^{2}} \delta_{m,0}.
\end{equation}
From previous considerations, we expect that superrotation zero-mode gives us the angular momentum of black holes. The equations \eqref{91} and \eqref{92} obey this property when the black hole does not have electric charge. But it is not true when electric charge turns on. To cure this problem, we consider two subalgebras of the algebra \eqref{71}. We construct them by introducing two new modes
\begin{equation}\label{93}
  \eta _{m} = \chi (0,0,0,-z^{m+1}), \hspace{1 cm} \bar{\eta} _{m} = \chi (0,0,0,-\bar{z}^{m+1}).
\end{equation}
One can show that $\eta _{m}$ and $\bar{\eta} _{m}$ obey the following commutation relations
\begin{equation}\label{94}
  \begin{split}
       & [Y_{m}, \eta_{n}] =-(n+1) \eta_{m+n}, \hspace{1 cm} [\bar{Y}_{m}, \bar{\eta}_{n}] =-(n+1) \bar{\eta}_{m+n}, \\
       & [\bar{Y}_{m}, \eta_{n}] =0 , \hspace{1 cm} [Y_{m}, \bar{\eta}_{n}] =0, \hspace{1 cm} [T_{(m,n)}, \eta_{k}]= 0, \hspace{1 cm} [T_{(m,n)}, \bar{\eta}_{k}] = 0,\\
       & [\hat{\lambda}_{(m,n)}, \eta_{k}] =0, \hspace{1 cm} [\hat{\lambda}_{(m,n)}, \bar{\eta}_{k}]=0, \hspace{1 cm} [\eta_{m}, \bar{\eta}_{n}] = 0,
  \end{split}
\end{equation}
which are extracted from the equations \eqref{65}-\eqref{68}. Now we define new superrotation modes as follows:
\begin{equation}\label{95}
  Y^{(\text{new})}_{m} = Y_{m} + \eta_{m}, \hspace{1 cm} \bar{Y}^{(\text{new})}_{m} = \bar{Y}_{m} + \bar{\eta}_{m},
\end{equation}
so that they obey same algebra as the old ones do. Strictly speaking, these new modes obey the algebra \eqref{70} and \eqref{71} with $Y_{m} \rightarrow Y^{(\text{new})}_{m}$ and $\bar{Y}_{m} \rightarrow \bar{Y}^{(\text{new})}_{m}$. Thus, we are allowed to use $U(1)$ gauge fixing to cure the problem. To this end, we fix the $U(1)$ gauge freedom as
\begin{equation}\label{96}
  \hat{\lambda}= Y^{A} \varphi_{A},
\end{equation}
so that corresponding modes are given by Eq.\eqref{93}. In this way, we can define new superrotation charges conjugate to superrotation modes  $Y^{(\text{new})}_{m}$ and $\bar{Y}^{(\text{new})}_{m}$ as
\begin{equation}\label{97}
  \mathcal{Y}^{(\text{new})}_{m}= \frac{1}{16 \pi} \int dz d \bar{z} \Omega \sqrt{\gamma} z^{m+1} \left( \theta_{z} + 16 \pi \psi_{v} \varphi_{z} \right),
\end{equation}
\begin{equation}\label{98}
  \bar{\mathcal{Y}}^{(\text{new})}_{m}= \frac{1}{16 \pi} \int dz d \bar{z} \Omega \sqrt{\gamma} \bar{z}^{m+1} \left( \theta_{\bar{z}} + 16 \pi \psi_{v} \varphi_{\bar{z}} \right),
\end{equation}
In fact, these are charge modes corresponding to charge conjugate to symmetry generator $\chi = \chi (0, Y^{z}, Y^{\bar{z}}, Y^{B} \varphi_{B})$. For the Kerr-Newman (A)dS black hole, we will have
\begin{equation}\label{99}
  \mathcal{Y}^{(\text{new})}_{m}=-i \frac{M a}{2 \Xi ^{2}} \delta_{m,0}, \hspace{1 cm} \bar{\mathcal{Y}}^{(\text{new})}_{m}=+i \frac{M a}{2 \Xi ^{2}} \delta_{m,0}.
\end{equation}
So the problem is cured. Since $\hat{\lambda}$ is in general a dynamical field independent function and we set it as Eq.\eqref{96} in the last step, i.e. when we want to calculate charges, then $\mathcal{Y}^{(\text{new})}_{m}$ and $\bar{\mathcal{Y}}^{(\text{new})}_{m}$ will satisfy the same algebra as \eqref{80} and \eqref{81}:
\begin{equation}\label{100}
  \begin{split}
       & \{ \mathcal{Y}^{(\text{new})}_{m}, \mathcal{Y}^{(\text{new})}_{n}\}= (m-n) \mathcal{Y}^{(\text{new})}_{m+n}, \hspace{1 cm} \{ \bar{\mathcal{Y}}^{(\text{new})}_{m}, \bar{\mathcal{Y}}^{(\text{new})}_{n}\}= (m-n) \bar{\mathcal{Y}}^{(\text{new})}_{m+n}, \\
       & \{ \mathcal{Y}^{(\text{new})}_{k}, \mathcal{T}_{(m,n)} \} = -m \mathcal{T}_{(m+k,n)}, \hspace{1 cm} \{ \bar{\mathcal{Y}}^{(\text{new})}_{k}, \mathcal{T}_{(m,n)} \}= -n \mathcal{T}_{(m,n+k)},\\
       &  \{\mathcal{Y}^{(\text{new})}_{k}, \mathcal{Q}_{(m,n)} \}= -m \mathcal{Q}_{(m+k,n)}, \hspace{1 cm} \{\bar{\mathcal{Y}}^{(\text{new})}_{k}, \mathcal{Q}_{(m,n)} \}= -n \mathcal{Q}_{(m,n+k)}.
  \end{split}
\end{equation}
where brackets not displayed vanish. We conclude this section by mentioning that $T_{(m,n)}$, $Y^{(\text{new})}_{m}$ and $\bar{Y}^{(\text{new})}_{m}$ are generators of soft hairs and $\hat{\lambda}_{(m,n)}$ are generators of soft electric hairs.
\section{Sugawara Deconstruction and New Entropy Formula}\label{S.VII}
Now, we define $\tilde{\mathcal{T}}_{(m,n)} = \frac{1}{2 \kappa} \mathcal{T}_{(m,n)}$ and replace the brackets with commutators, namely $ \{ \hspace{2 mm},\hspace{2 mm} \} \equiv i [\hspace{2 mm},\hspace{2 mm}]$, then \eqref{100} becomes
\begin{equation}\label{101}
  \begin{split}
       & i [ \mathcal{Y}^{(\text{new})}_{m}, \mathcal{Y}^{(\text{new})}_{n}]= (m-n) \mathcal{Y}^{(\text{new})}_{m+n}, \hspace{1 cm} i [ \bar{\mathcal{Y}}^{(\text{new})}_{m}, \bar{\mathcal{Y}}^{(\text{new})}_{n}]= (m-n) \bar{\mathcal{Y}}^{(\text{new})}_{m+n}, \\
       & i [ \mathcal{Y}^{(\text{new})}_{k}, \tilde{\mathcal{T}}_{(m,n)} ] = - m \tilde{\mathcal{T}}_{(m+k,n)}, \hspace{1 cm} i [ \bar{\mathcal{Y}}^{(\text{new})}_{k}, \tilde{\mathcal{T}}_{(m,n)} ]= - n \tilde{\mathcal{T}}_{(m,n+k)},\\
       & i [ \mathcal{Y}^{(\text{new})}_{k}, \mathcal{Q}_{(m,n)} ]= - m \mathcal{Q}_{(m+k,n)}, \hspace{1 cm} i [\bar{\mathcal{Y}}^{(\text{new})}_{k}, \mathcal{Q}_{(m,n)} ]= - n \mathcal{Q}_{(m,n+k)}.
  \end{split}
\end{equation}
where commutators not displayed vanish. It is clear form Eq.\eqref{85} and Eq.\eqref{86} that the supertranslation double-zero-mode $\tilde{\mathcal{T}}_{(0,0)}$ is related to the Kerr-Newman (A)dS black hole entropy as
\begin{equation}\label{102}
  S= 4 \pi \tilde{\mathcal{T}}_{(0,0)}.
\end{equation}
Therefore, we can apply the Sugawara deconstruction proposed in \cite{18}. To do this, we introduce four new generators $\hat{J}^{\pm}_{m}$ and $\hat{K}^{\pm}_{m}$ so that they obey the following algebra
\begin{equation}\label{103}
  i [\hat{J}^{\pm}_{m}, \hat{K}^{\pm}_{n}]= m \delta_{m+n,0},
\end{equation}
where commutators not displayed vanish. The algebra \eqref{103} consists of two copies of the 3-dimensional flat space near horizon symmetry algebra \cite{19}. Hence we can construct generators $\tilde{\mathcal{T}}_{(m,n)}$ , $\mathcal{Y}^{(\text{new})}_{m}$ and $\bar{\mathcal{Y}}^{(\text{new})}_{m}$ as follows:
\begin{equation}\label{104}
      \tilde{\mathcal{T}}_{(m,n)} = \hat{J}^{+}_{m} \hat{J}^{-}_{n}, \hspace{1 cm}  \mathcal{Y}^{(\text{new})}_{m} = \sum_{p} \hat{J}^{+}_{m-p} \hat{K}^{+}_{p}, \hspace{1 cm} \bar{\mathcal{Y}}^{(\text{new})}_{m} = \sum_{p} \hat{J}^{-}_{m-p} \hat{K}^{-}_{p}.
\end{equation}
It is easy to check that the definitions presented in Eq.\eqref{104} obey the algebra \eqref{101} provided that $\hat{J}^{\pm}_{m}$ and $\hat{K}^{\pm}_{m}$ satisfy the algebra introduced in Eq.\eqref{103}. By comparing Eq.\eqref{104} and equations \eqref{89},\eqref{99} and \eqref{102}, we find that there exist six algebraic constraints on zero modes (because we assume that zero modes are complex numbers)
\begin{equation}\label{105}
 \frac{S}{4 \pi} = \hat{J}^{+}_{0} \hat{J}^{-}_{0},\hspace{1 cm}  -\frac{i}{2} \mathcal{J} = \hat{J}^{+}_{0} \hat{K}^{+}_{0}, \hspace{1 cm} + \frac{i}{2} \mathcal{J} = \hat{J}^{-}_{0} \hat{K}^{-}_{0},
\end{equation}
where Eq.\eqref{35} was used. The expression for the angular momentum, introduced in Eq.\eqref{105}, could be a linear combination as $ \mp \frac{i}{2} \mathcal{J} =\sum_{p} \hat{J}^{\pm}_{-p} \hat{K}^{\pm}_{p}$. However, we assume that just the zero modes of $\hat{K}^{\pm}_{m}$ are non-zero, i.e. $\hat{K}^{\pm}_{m}$ are proportional to kronecker delta $\delta _{m,0}$, which is in agreement with Eq.\eqref{99}. In order to determine zero modes uniquely, we need two other constraints. Two constrains were presented in \cite{18}. The authors in \cite{18} have claimed that we should not see the angular momentum in chirally symmetric sum of zero modes. Also, they demand that chirally
symmetric sum of zero modes to be equal to two times black hole mass. But it seems that this constrain is no longer held in presence of cosmological constant and/or when black hole has electric charge. One of the choices that can be made is
\begin{equation}\label{106}
  \hat{J}^{\pm}_{0}= \frac{1}{2 \Xi^{\frac{1}{2}}} (r_{+} \mp i a), \hspace{1 cm} \hat{K}^{\pm}_{0}= \frac{M a}{\Xi^{\frac{3}{2}} (r_{+}^{2}+a^{2})} (a \mp i r_{+}),
\end{equation}
which will reduce to the ones proposed in \cite{18} when we set $\Lambda=0$ and $Q=0$. One can show that $ \hat{K}^{+}_{0} \hat{K}^{-}_{0}$ is given by
\begin{equation}\label{108}
  \hat{K}^{+}_{0} \hat{K}^{-}_{0}= \frac{M^{2} a^{2}}{\Xi^{3} (r_{+}^{2}+a^{2}) },
\end{equation}
and it does not provide inner horizon entropy of the Kerr-Newman (A)dS black hole (however, for Kerr black hole it does). This result is independent of the choice \eqref{106}. The chirally symmetric sum of zero modes is
\begin{equation}\label{109}
  \hat{J}^{+}_{0}+ \hat{J}^{-}_{0} + \hat{K}^{+}_{0}+ \hat{K}^{-}_{0}= \frac{2 \Xi^{\frac{1}{2}}}{\left(1- \frac{\Lambda}{3}r_{+}^{2} \right)}\left[ \mathcal{M} - \frac{1}{2} \mathcal{Q} \Phi_{E}\right],
\end{equation}
where
\begin{equation}\label{110}
  \Phi_{E} = \frac{Q}{4 \pi} \frac{r_{+}}{(r_{+}^{2}+a^{2})},
\end{equation}
is the horizon electric potential. Thus, we have shown that the Kerr-Newman (A)dS black hole entropy is bilinear in the zero modes and it satisfy the new entropy formula $ S= 4 \pi\hat{J}^{+}_{0} \hat{J}^{-}_{0}$ proposed in \cite{18}. It is also clear from Eq.\eqref{105} that the angular momentum is given by $ \mathcal{J}=i (\hat{J}^{+}_{0} \hat{K}^{+}_{0} - \hat{J}^{-}_{0} \hat{K}^{-}_{0})$.
\section{Conclusion}
We have briefly reviewed the approach of the covariant phase space method of obtaining conserved charges in Einstein-Maxwell theory. According to \cite{6}, we introduced combined symmetry generator $\chi=(\xi, \lambda)$, which consists of diffeomorphism and $U(1)$ gauge transformation. The covariant phase space method presented in section \ref{S.II}, is not only applicable to the asymptotic symmetries at spatial infinity but also it is applicable to the near horizon asymptotic symmetries. In section \ref{S.III}, we have briefly reviewed the Kerr-Newman (A)dS black hole geometry and then we found the near horizon behavior of this black hole in the Gaussian null coordinate system. We showed that the induced metric on the horizon is conformally related to the Riemann sphere. Therefore the Kerr-Newman (A)dS black hole horizon admits conformal symmetry. The explicit form of conformal factor $\Omega= \Omega(z,\bar{z})$ depends on the sign of cosmological constant (see Eq.\eqref{54} and Eq.\eqref{56}). Therefore, we expect that fall-off conditions near the isolated horizon in the Einstein-Maxwell theory are given by Eq.\eqref{57} and Eq\eqref{58}, where $\kappa$ and $\varphi_{v}$ are constants (imposed by field equations). These fall-off conditions are preserved by the action of asymptotic symmetry generators $\chi$ (components of $\chi$ are given by Eq.\eqref{60} and Eq.\eqref{61}). Equations \eqref{65}-\eqref{68} give the algebra among these asymptotic symmetry generators. Because the algebra among the conserved charges and the asymptotic symmetry generators must be isomorphic, the additional commutation relation \eqref{68} was introduced. The asymptotic symmetry generator modes, satisfy an algebra contains of a set of supertranslations current $T_{(m,n)}$, two sets of the Witt algebra currents, given by $Y_{m}$ and $\bar{Y}_{m}$, and a set of multiple-charges current $\hat{\lambda}_{(m,n)}$. Two sets of the Witt currents are in semi-direct sum with the supertranslations and multiple-charges current. In section \ref{S.VI}, we found conserved charges conjugate to these modes. The supertranslation double-zero-mode charge $\mathcal{T}_{(0,0)}$ gives the Kerr-Newman (A)dS black hole entropy multiplied by Hawking temperature. Also, the multiple-charge double-zero-mode gives the electric charge of the Kerr-Newman (A)dS black hole. One expect that the zero-mode charges associated to superrotations give angular momentum of black holes. But it is not true when black hole have electric charge (See equations \eqref{91} and \eqref{92}). Due to the presence of second terms in the right hand sides of Eq.\eqref{91} and \eqref{92}, this problem occurs when one considers just diffeomorphism generated by $Y^{A}$. To cure this problem, we must consider both diffeomorphism and $U(1)$ gauge transformation together. To do this, we introduced a transformation generated by $\chi (Y)= \chi(0, Y^{A}, \hat{\lambda}(Y)) \rightarrow Y$, where $\hat{\lambda}$ is a function of $Y^{A}$ and it does not generate an independent symmetry. We defined corresponding zero modes as $Y^{\text{(new)}}_{m}$ and $\bar{Y}^{\text{(new)}}_{m}$. By a gauge fixing as $\hat{\lambda}= Y^{A} \varphi_{A}$, we have found charges conjugate to these modes. The zero mode charges give the angular momentum of the Kerr-Newman (A)dS black hole (see Eq.\eqref{99}). A question remains still open: it remains to be understood the physical meaning of such a modification of charges associated to supperrotations. We showed that the algebra among charge modes are given by \eqref{101}. We used the Sugawara deconstruction to show that the new entropy formula proposed in \cite{18} is held for the Kerr-Newman (A)dS black hole.

\section{Acknowledgments}
The work of Hamed Adami has been financially supported by Research Institute for Astronomy Astrophysics of Maragha (RIAAM).


\begin{thebibliography}{References}
\bibitem{1'}J. D. Bekenstein, Lett. Nuovo. Cim. 4, 737 (1972); Phys. Rev. D7, 2333 (1973); Phys.
Rev. D9, 3292 (1974).
\bibitem{2'}S. W. Hawking, Comm. Math. Phys. 25, 152 (1972); J. M. Bardeen, B. Carter and
S. W. Hawking, Comm. Math. Phys. 31, 161 (1973).
\bibitem{26a} S. W. Hawking, M. J. Perry, A. Strominger, Phys. Rev. Lett. \textbf{116} (2016) 231301.
\bibitem{6'} S. W. Hawking, M. J. Perry, A. Strominger, JHEP \textbf{1705} (2017) 161.
\bibitem{7'} A. Strominger, arXiv:1703.05448 [hep-th].
\bibitem{19} H. Afshar, D. Grumiller, W. Merbis, A. Perez, D. Tempo, R. Troncoso, Phys. Rev. D \textbf{95} (2017) 106005.
\bibitem{3'}M. R. Setare, H. Adami, arXiv:1711.08344[hep-th].
\bibitem{4'}D. Grumiller, P. Hacker, W. Merbis,	arXiv:1711.07975 [hep-th].
\bibitem{140}M. R. Setare, Nucl. Phys. B 898, 259, (2015).
\bibitem{5'} M. R. Setare , H. Adami, Nucl. Phys. B 914, 220, (2017).
\bibitem{18} H. Gonzalez, D. Grumiller, W. Merbis, R. Wutte, EPJ Web Conf. \textbf{168} (2018) 01009.
\bibitem{1} J. Lee and R. M. Wald, J. Math. Phys. \textbf{31} (1990) 725.
\bibitem{2} R. M. Wald, Phys. Rev. D \textbf{48} (1993) 3427.
\bibitem{3} V. Iyer and R. M. Wald, Phys. Rev. D \textbf{50} (1994) 846.
\bibitem{4} K. Prabhu, Class. Quant. Grav. \textbf{34} (2017), 035011.
\bibitem{5} K. Hajian, M. M. Sheikh-Jabbari, Phys. Rev. D \textbf{93} (2016) 044074.
\bibitem{6} M. R. Setare, H. Adami, Class. Quant. Grav. \textbf{34} (2017), 105008.
\bibitem{7} M. M. Caldarelli, G. Cognola, D. Klemm, Class. Quant. Grav. \textbf{17} (2000) 399.
\bibitem{8} K. Hajian, Gen. Rel. Grav. \textbf{48} (2016), 114.
\bibitem{9} I. Booth, Phys. Rev. D \textbf{87} (2013), 024008.
\bibitem{10} L. A. Tamburino and J. H. Winicour, Phys. Rev. \textbf{150} (1966) 1039.
\bibitem{11} G. Barnich and C. Troessaert, JHEP \textbf{1005} (2010) 062.
\bibitem{12} L. Donnay, G. Giribet, H. A. Gonzalez and M. Pino, JHEP \textbf{1609} (2016) 100.
\bibitem{16} L. Donnay, G. Giribet, H. A. Gonzalez, M. Pino, Phys. Rev. Lett. \textbf{116} (2016), 091101.
\bibitem{13} G. Barnich and C. Troessaert, JHEP \textbf{1005} (2010) 062.
\bibitem{14} G. Barnich and C. Troessaert, Phys. Rev. Lett. \textbf{105} (2010) 111103.
\bibitem{15} G. Barnich and C. Troessaert, JHEP \textbf{1112} (2011) 105.
\bibitem{17} M. R. Setare, H. Adami, Phys. Lett. B \textbf{760} (2016) 411.
\end{thebibliography}
\end{document}